\documentclass[runningheads]{llncs}
\usepackage{accvabbrv}
\usepackage{graphicx}
\usepackage{multirow}
\usepackage{amsmath,amssymb}
\usepackage[table]{xcolor}
\usepackage{booktabs}
\usepackage{array}
\usepackage{makecell}
\usepackage{hyperref}
\usepackage[accsupp]{axessibility}
\usepackage{capt-of}
\usepackage{tikz}
\usetikzlibrary{arrows.meta,positioning,matrix,calc}

\tikzset{
    erbox/.style={draw, rounded corners=2pt, align=center, minimum height=8mm, text width=23mm, font=\scriptsize},
    erwidebox/.style={draw, rounded corners=2pt, align=center, minimum height=9mm, text width=31mm, font=\scriptsize},
    erarrow/.style={-{Latex[length=1.5mm]}, thick},
    erline/.style={thick},
    erloss/.style={draw, rounded corners=2pt, align=center, fill=orange!12, minimum height=9mm, text width=28mm, font=\scriptsize},
    eranchor/.style={draw, rounded corners=2pt, align=center, fill=blue!8, minimum height=9mm, text width=29mm, font=\scriptsize}
}

\makeatletter
\renewcommand\subsection{\@startsection{subsection}{2}{\z@}%
  {-10\p@ \@plus -2\p@ \@minus -2\p@}%
  {4\p@ \@plus 2\p@ \@minus 2\p@}%
  {\normalfont\normalsize\bfseries\boldmath
   \rightskip=\z@ \@plus 8em\pretolerance=10000 }}
\makeatother

\begin{document}

\title{ERCLIP: Small-Data Adaptation of CLIP for Free-Length Image-Text Retrieval}
\titlerunning{ERCLIP: Small-Data Adaptation of CLIP}
\author{Hoang-Bao Le$^1$, Allie Tran$^1$, Binh T. Nguyen$^2$, Liting Zhou$^1$, Cathal Gurrin$^1$}
\authorrunning{Le et al.}

\institute{ADAPT Centre, School of Computing, Dublin City University, Dublin, Ireland \and
Ho Chi Minh University of Science, Vietnam National University, Ho Chi Minh City, Vietnam\\
\email{bao.le2@mail.dcu.ie, \{allie.tran, liting.zhou, cathal.gurrin\}@dcu.ie}}

\maketitle
\vspace{-1.5em}

\begin{abstract}
CLIP's 77-token text encoder truncates detailed image descriptions. We present ERCLIP, a small-data adaptation of CLIP for long-caption image--text retrieval. During training, a long caption is split into sentences that the CLIP text encoder encodes independently; the sentence embeddings are fused by mean pooling or a gated residual Transformer scorer and receive their own contrastive loss. At inference ERCLIP is an unmodified CLIP dual encoder: test captions are still read through the 77-token window, and only training sees the full caption. We find that this fine-tuning erodes CLIP's zero-shot classification, most when the fusion output is not a combination of CLIP embeddings, and add a frozen-CLIP preservation loss on the image and short-caption embeddings. With 20k pairs and one epoch, ERCLIP improves long-caption R@1 over CLIP by 6.5--11.2 points, closing 62--77\% of the gap to LongCLIP with 2\% of its long captions. Preservation recovers 1.5--2.1 ImageNet zero-shot points (30--42\% of the loss) and adds up to 2.4 points of COCO image-to-text R@1, at no inference cost.
\end{abstract}

\section{Introduction}
CLIP dual encoders align images and text in a shared embedding space and support both retrieval and zero-shot recognition \cite{radford2021learning}. Their text encoders, however, read at most 77 tokens, so detailed descriptions are truncated and the discarded facts never reach the embedding. Long-context methods extend the positional embeddings, match features across caption lengths, or learn token selection \cite{zhang2024longclip,xie2025smartclip,najdenkoska2025tulip,wu2026himoclip}. They rely on large long-caption corpora and rarely report what the adaptation does to the tasks CLIP already solves, a cost documented for classification fine-tuning \cite{wortsman2022robust,kumar2022finetuning} but not for long-caption retrieval.

We study the opposite regime: 20k image--caption pairs, 2\% of the long captions used by LongCLIP, and one epoch on one GPU. ERCLIP splits each long caption into sentences that fit the original encoder, fuses their embeddings, and aligns the result with the image. The sentence pathway exists only in training, so the adapted model keeps CLIP's architecture, 77-token context, and inference cost; at test time long captions are still truncated, but the part the encoder reads is better aligned with the image. We observe that this adaptation erodes zero-shot classification by an amount that depends on the fusion module and grows with training, and we limit it with a gated residual fusion initialised at mean pooling and a frozen-CLIP preservation loss.

Our contributions are (i) a sentence-level scheme that adapts CLIP to long captions from 20k pairs without changing its inference; (ii) evidence that the fusion module sets the cost to zero-shot ability, which motivates gated residual fusion and a label-free preservation loss; and (iii) a joint evaluation of long-caption retrieval, short-caption retrieval, and zero-shot classification at one fixed budget. The ingredients are not new (Sec.~\ref{sec:related}); our focus is their behaviour under scarce data.

\section{Related Work}
\label{sec:related}
\noindent\textbf{Vision--language pretraining.} CLIP learns image and text encoders contrastively \cite{radford2021learning}; follow-up work changes the data, objective, or supervision \cite{jia2021align,yang2023alip,li2023declip,zhai2023sigmoid,li2022blip,li2023blip2}, all at a scale far beyond our budget.

\noindent\textbf{Long-caption retrieval.} LongCLIP stretches the positional embeddings to 248 tokens and aligns the principal components of short-caption features \cite{zhang2024longclip}. SmartCLIP and TULIP learn structured selection or relative positional encodings for long text \cite{xie2025smartclip,najdenkoska2025tulip}, and HiMo-CLIP models semantic hierarchy and monotonicity \cite{wu2026himoclip}. All are trained on large corpora, e.g., 1M ShareGPT4V captions for LongCLIP. LongCLIP's ablations include aligning an image with both its long and short caption, and anchoring short-text features to the frozen CLIP encoder, and find both inferior to its own strategy \cite{zhang2024longclip}. ERCLIP's objective is close to these variants, and we do not claim the losses as new. We study them in a different setting: a 77-token encoder that reaches content beyond its window only through sentence-level fusion during training, an anchor on the image as well as the short caption, and 20k pairs for one epoch. We ask how much of the full-scale gain this recovers, and at what cost, without large data or compute.

\noindent\textbf{Preserving pretrained behaviour.} Fine-tuning CLIP degrades its zero-shot ability. WiSE-FT recovers much of it by weight interpolation \cite{wortsman2022robust}, LP-FT shows that full fine-tuning distorts pretrained features \cite{kumar2022finetuning}, and FLYP retains more by fine-tuning with the contrastive pretraining loss \cite{goyal2023finetune}; continual learning earlier used distillation to a frozen copy \cite{li2018learning} or parameter regularisation \cite{kirkpatrick2017overcoming}. These works fine-tune for and preserve classification; we fine-tune for long-caption retrieval, a different task.

\section{Method}
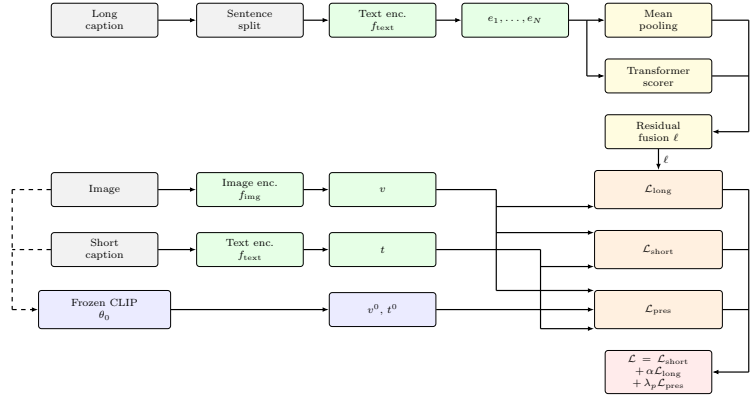
\begin{figure}[t]
\centering
\resizebox{.8\textwidth}{!}{%
\begin{tikzpicture}[font=\small]

\matrix (m) [matrix, column sep=6mm, row sep=5mm, ampersand replacement=\&] {
\node[erbox, fill=gray!10]  (long)  {Long\\caption};                  \&
\node[erbox, fill=gray!10]  (split) {Sentence\\split};                \&
\node[erbox, fill=green!10] (ltext) {Text enc.\\$f_{\mathrm{text}}$}; \&
\node[erbox, fill=green!10] (sent)  {$e_1,\ldots,e_N$};               \&
\node[erbox, fill=yellow!15](mean)  {Mean\\pooling};                  \\
\& \& \& \&
\node[erbox, fill=yellow!15](scorer){Transformer\\scorer};            \\
\& \& \& \&
\node[erbox, fill=yellow!15](residual){Residual\\fusion $\ell$};      \\
\node[erbox, fill=gray!10]  (img)   {Image};                          \&
\node[erbox, fill=green!10] (venc)  {Image enc.\\$f_{\mathrm{img}}$}; \&
\node[erbox, fill=green!10] (v)     {$v$};                            \& \&
\node[erloss]               (llong) {$\mathcal{L}_{\mathrm{long}}$};  \\
\node[erbox, fill=gray!10]  (short) {Short\\caption};                 \&
\node[erbox, fill=green!10] (stext) {Text enc.\\$f_{\mathrm{text}}$}; \&
\node[erbox, fill=green!10] (t)     {$t$};                            \& \&
\node[erloss]               (lshort){$\mathcal{L}_{\mathrm{short}}$}; \\
\node[eranchor]             (teacher){Frozen CLIP\\$\theta_0$};       \& \&
\node[erbox, fill=blue!8]   (anchor){$v^{0},\,t^{0}$};                \& \&
\node[erloss]               (lpres) {$\mathcal{L}_{\mathrm{pres}}$};  \\
\& \& \& \&
\node[erbox, fill=red!8]    (total) {$\mathcal{L}=\mathcal{L}_{\mathrm{short}}$\\%
                                     $+\,\alpha\mathcal{L}_{\mathrm{long}}$\\%
                                     $+\,\lambda_p\mathcal{L}_{\mathrm{pres}}$}; \\
};

\coordinate (bA) at ($(sent.east)!0.5!(mean.west)$);
\coordinate (bB) at ([xshift= 5mm]m.east);
\coordinate (bv) at ($(v.east)!0.38!(llong.west)$);
\coordinate (bt) at ($(v.east)!0.66!(llong.west)$);
\coordinate (bl) at ([xshift=-5mm]m.west);

\coordinate (a1) at ([yshift=-4.0mm]llong.west);
\coordinate (a2) at ([yshift= 4.0mm]lshort.west);
\coordinate (a3) at ([yshift= 4.5mm]lpres.west);
\coordinate (a4) at ([yshift=-4.0mm]lshort.west);
\coordinate (a5) at ([yshift=-4.5mm]lpres.west);

\draw[erarrow] (long)  -- (split);
\draw[erarrow] (split) -- (ltext);
\draw[erarrow] (ltext) -- (sent);

\draw[erline]  (sent.east)    -- (bA |- sent);
\draw[erline]  (bA |- mean)   -- (bA |- scorer);
\draw[erarrow] (bA |- mean)   -- (mean.west);
\draw[erarrow] (bA |- scorer) -- (scorer.west);

\draw[erline]  (mean.east)      -- (bB |- mean);
\draw[erline]  (scorer.east)    -- (bB |- scorer);
\draw[erline]  (bB |- mean)     -- (bB |- residual);
\draw[erarrow] (bB |- residual) -- (residual.east);

\draw[erarrow] (residual.south) -- node[right,font=\scriptsize]{$\ell$} (llong.north);

\draw[erarrow] (img)   -- (venc);
\draw[erarrow] (venc)  -- (v);
\draw[erarrow] (short) -- (stext);
\draw[erarrow] (stext) -- (t);

\draw[erline]  (v.east)   -- (bv |- v);     
\draw[erline]  (bv |- v)  -- (bv |- a3);
\draw[erarrow] (bv |- a1) -- (a1);
\draw[erarrow] (bv |- a2) -- (a2);
\draw[erarrow] (bv |- a3) -- (a3);

\draw[erline]  (t.east)   -- (bt |- t);     
\draw[erline]  (bt |- t)  -- (bt |- a5);
\draw[erarrow] (bt |- a4) -- (a4);
\draw[erarrow] (bt |- a5) -- (a5);

\draw[erline,dashed]  (img.west)      -- (bl |- img);
\draw[erline,dashed]  (short.west)    -- (bl |- short);
\draw[erline,dashed]  (bl |- img)     -- (bl |- teacher);
\draw[erarrow,dashed] (bl |- teacher) -- (teacher.west);
\draw[erarrow] (teacher)     -- (anchor);
\draw[erarrow] (anchor.east) -- (lpres.west);

\draw[erline]  (llong.east)  -- (bB |- llong);
\draw[erline]  (lshort.east) -- (bB |- lshort);
\draw[erline]  (lpres.east)  -- (bB |- lpres);
\draw[erline]  (bB |- llong) -- (bB |- total);
\draw[erarrow] (bB |- total) -- (total.east);
\end{tikzpicture}}
\caption{ERCLIP training graph. Green encoders are trainable; the two text encoders
share weights. Sentences $s_1,\ldots,s_N$ are encoded into $e_1,\ldots,e_N$ and fused
as $\ell=z_{\mathrm{mean}}+\gamma(z_{\mathrm{attn}}-z_{\mathrm{mean}})$, with weights
from the Transformer scorer (yellow). A frozen copy $\theta_0$ supplies $v^{0},t^{0}$
for $\mathcal{L}_{\mathrm{pres}}$. \emph{At inference only the green encoders remain}:
images pass through $f_{\mathrm{img}}$ and captions, truncated to 77 tokens, through
$f_{\mathrm{text}}$; sentence split, fusion, teacher, and losses (orange) are discarded.}
\label{fig:framework}
\vspace{-1em}
\end{figure}

ERCLIP fine-tunes the CLIP encoders $f_{\mathrm{img}}$ and $f_{\mathrm{text}}$ (embedding dimension $d$) on triplets of an image $x$, a short caption $c$, and a long caption $L$ (Fig.~\ref{fig:framework}).

\subsection{Sentence-Level Encoding and Fusion}
$L$ is split on sentence-final periods into $N$ ordered sentences, each tokenised separately and encoded by the shared text encoder, $e_i=f_{\mathrm{text}}(s_i)\in\mathbb{R}^d$, so that every sentence fits the pretrained encoder's window. The embeddings $E=[e_1;\ldots;e_N]$ are fused into one long-caption embedding.

\noindent\textbf{Mean pooling} $z_{\mathrm{mean}}=\frac{1}{N}\sum_i e_i$ adds no parameters and weights every sentence equally.

\noindent\textbf{Transformer scorer.} A single T5-style Transformer block \cite{raffel2020t5} (8 heads) contextualises $E+P$, where $P_i$ is the $i$-th row of CLIP's text positional embedding and encodes sentence order. An MLP $g$ scores each contextual state $h_i$:
\begin{equation}
 H=\mathrm{TBlock}(E+P),\qquad w_i=\operatorname{softmax}_i\,g(h_i),\qquad z_{\mathrm{attn}}=\textstyle\sum_i w_i e_i .
\end{equation}
The scorer re-weights the original $e_i$ rather than emitting $h_i$, so $z_{\mathrm{attn}}$ stays inside the convex hull of CLIP sentence embeddings.

\noindent\textbf{Gated residual fusion.} ERCLIP-v1 uses $z_{\mathrm{attn}}$ directly, so its randomly initialised scorer co-adapts with the backbone from arbitrary weights. We instead start from mean pooling with a scalar gate $\gamma$ initialised to zero:
\begin{equation}
 \ell=\frac{z_{\mathrm{long}}}{\|z_{\mathrm{long}}\|_2},\qquad z_{\mathrm{long}}=z_{\mathrm{mean}}+\gamma\,(z_{\mathrm{attn}}-z_{\mathrm{mean}}).
\end{equation}
The model starts as mean pooling, and the scorer acts only as far as $\gamma$ grows. As an ablation, v1 also uses an LSTM \cite{hochreiter1997lstm} whose final hidden state is $z_{\mathrm{long}}$, an output not formed from CLIP embeddings.

\subsection{Training Objective}
Let $v_b$, $t_b$, and $\ell_b$ be the $\ell_2$-normalised image, short-caption, and fused long-caption embeddings of a batch of size $B$. We use the symmetric InfoNCE loss with a fixed temperature $\tau=0.07$ and label smoothing $\epsilon=0.1$,
\begin{equation}
 \mathcal{L}_{\mathrm{clip}}(V,Y)=\tfrac{1}{2}\big[\operatorname{CE}_{\epsilon}(VY^\top/\tau,\,I)+\operatorname{CE}_{\epsilon}(YV^\top/\tau,\,I)\big],
\end{equation}
with $\mathcal{L}_{\mathrm{short}}=\mathcal{L}_{\mathrm{clip}}(V,T)$ and $\mathcal{L}_{\mathrm{long}}=\mathcal{L}_{\mathrm{clip}}(V,\Lambda)$: two in-batch contrastive terms sharing the image embedding, not a triplet loss. Since every sentence passes through the $f_{\mathrm{text}}$ that also encodes short captions and class prompts, the long-caption gradient moves the representation zero-shot classification relies on. ERCLIP-v1 optimises only $\mathcal{L}_{\mathrm{short}}+\alpha\mathcal{L}_{\mathrm{long}}$, which costs 3--6 ImageNet points with Transformer fusion and 12--14 with the LSTM (Sec.~\ref{sec:basic}).

\noindent\textbf{Frozen-CLIP preservation.} Before training we copy the model into a frozen teacher $\theta_0$ \cite{li2018learning}, which gives reference embeddings $v^0_b,t^0_b$ for the same image and short caption, and penalise the cosine distance to them:
\begin{equation}
 \mathcal{L}_{\mathrm{pres}}=\frac{1}{2B}\sum_{b=1}^{B}\big[(1-v_b^{\top}v^0_b)+(1-t_b^{\top}t^0_b)\big],\qquad
 \mathcal{L}=\mathcal{L}_{\mathrm{short}}+\alpha\,\mathcal{L}_{\mathrm{long}}+\lambda_p\,\mathcal{L}_{\mathrm{pres}}.
\end{equation}
Only the image and short-caption embeddings, the interface of zero-shot classification, are anchored. The long-caption embedding stays free, since pretrained CLIP has no good long-caption representation to anchor it to (Tab.~\ref{tab:long_text_retrieval}). Unlike weight interpolation after training \cite{wortsman2022robust}, $\mathcal{L}_{\mathrm{pres}}$ anchors features during training and needs no labels or extra data.

\noindent\textbf{Inference.} Fusion and teacher are used only in training. At test time ERCLIP is the adapted dual encoder with CLIP's 77-token context, so longer captions are truncated exactly as in CLIP. ``Free-length'' thus refers to training, where every sentence reaches the encoder; the gains in Tab.~\ref{tab:long_text_retrieval} come from better alignment of the part of the caption the encoder reads, not from reading all of it. Sentence-wise encoding at test time is left to future work. ERCLIP keeps the parameters and latency of its backbone.

\section{Experiments}
\noindent\textbf{Data and metrics.} We train on the first 20k pairs, in COCO image-id order, of the 50k-image pool shared by COCO and ShareGPT4V \cite{lin2015microsoftcococommonobjects,chen2024sharegpt4v}, with the first COCO caption as the short caption and the ShareGPT4V description as the long one. We report Recall@1 (R@1) for image-to-text (I2T) and text-to-image (T2I) retrieval on Urban1K, ShareGPT4V-1k, DOCCI, and Long-DCI \cite{zhang2024longclip,onoe2024docci,Urbanek_2024_CVPR,najdenkoska2025tulip} and on the 5k images of COCO val2017, and zero-shot accuracy on CIFAR-10/100 and ImageNet \cite{krizhevsky2009cifar,russakovsky2015imagenet} with the CLIP prompt ensemble. Only Urban1K overlaps with our training data: 61 of its images are in our training set and 170 in our pool. For comparability with published results we report the full Urban1K set; on the 830 clean images every run's gain over CLIP changes by at most 1.2 points and the ranking of the variants is unchanged.

\noindent\textbf{Implementation.} We fine-tune all parameters of the ViT-B/32, B/16, and L/14 CLIP backbones with AdamW \cite{loshchilov2017adamw} (learning rate $5\times10^{-6}$, weight decay 0.05, 200 warm-up steps, cosine decay), batch size 64, mixed precision, and $\alpha=\lambda_p=1$ on one A100 GPU (312 steps per epoch). We compare M (mean pooling, $\lambda_p=0$), M+P (mean pooling, $\lambda_p=1$), and RT+P (residual Transformer, $\lambda_p=1$). Every run is evaluated after each of five epochs, and all tables report \emph{epoch 1}, a budget fixed in advance (Sec.~\ref{sec:length}). Each configuration is a single run, so we treat differences below about one point between variants as ties and draw conclusions only from effects with the same sign on all three backbones.

\begin{table}[t]
\centering
\setlength{\tabcolsep}{3pt}
\resizebox{.8\linewidth}{!}{%
\begin{tabular}{ll rr rr rr rr rr}
\toprule
\textbf{Model} & \textbf{ViT} & \multicolumn{2}{c}{\textbf{Urban1K}} & \multicolumn{2}{c}{\textbf{ShareGPT4V}} & \multicolumn{2}{c}{\textbf{DOCCI}} & \multicolumn{2}{c}{\textbf{Long-DCI}} & \multicolumn{2}{c}{\textbf{Avg.}} \\
\cmidrule(lr){3-4}\cmidrule(lr){5-6}\cmidrule(lr){7-8}\cmidrule(lr){9-10}\cmidrule(lr){11-12}
 & & I2T & T2I & I2T & T2I & I2T & T2I & I2T & T2I & I2T & T2I \\
\midrule
\textcolor{gray}{LongCLIP} & B/16 & 79.30 & 79.40 & 94.60 & 93.40 & 63.18 & 71.50 & 42.20 & 48.46 & 69.82 & 73.19 \\
\textcolor{gray}{LongCLIP} & L/14 & 82.50 & 86.10 & 97.20 & 97.30 & 66.58 & 78.56 & 46.53 & 54.28 & 73.20 & 79.06 \\
\textcolor{gray}{SmartCLIP} & B/16 & 90.50 & 87.20 & 98.70 & 98.00 & 77.38 & 78.08 & 54.80 & 54.49 & 80.34 & 79.44 \\
\textcolor{gray}{SmartCLIP} & L/14 & 93.30 & 90.10 & 97.90 & 98.50 & 81.70 & 82.58 & 59.14 & 60.27 & 83.01 & 82.86 \\
\textcolor{gray}{TULIP} & B/16 & 88.10 & 86.60 & 98.60 & 98.60 & -- & -- & 50.20 & 50.60 & -- & -- \\
\textcolor{gray}{TULIP} & L/14 & 90.10 & 91.10 & 99.00 & 99.00 & -- & -- & 55.70 & 56.40 & -- & -- \\
\midrule
CLIP & B/32 & 61.20 & 46.60 & 83.30 & 77.10 & 58.04 & 53.40 & 34.79 & 30.60 & 59.33 & 51.93 \\
ERCLIP M & B/32 & 61.40 & 65.40 & 83.70 & 81.00 & 58.66 & 60.24 & 38.54 & 39.88 & 60.57 & 61.63 \\
\rowcolor{gray!12}ERCLIP M+P & B/32 & 62.30 & 66.40 & 84.00 & 82.60 & 59.58 & 61.58 & 39.65 & 40.94 & 61.38 & 62.88 \\
\rowcolor{gray!12}ERCLIP RT+P & B/32 & 62.50 & 65.40 & 84.70 & 82.00 & 58.78 & 61.22 & 38.92 & 40.61 & 61.22 & 62.31 \\
\midrule
CLIP & B/16 & 67.50 & 53.30 & 84.50 & 79.60 & 60.76 & 57.10 & 36.91 & 33.73 & 62.42 & 55.93 \\
ERCLIP M & B/16 & 69.30 & 75.10 & 86.30 & 83.10 & 63.20 & 67.48 & 41.86 & 44.76 & 65.17 & 67.61 \\
\rowcolor{gray!12}ERCLIP M+P & B/16 & 68.60 & 75.70 & 87.50 & 85.30 & 63.24 & 67.16 & 41.57 & 45.34 & 65.23 & 68.38 \\
\rowcolor{gray!12}ERCLIP RT+P & B/16 & 69.30 & 73.40 & 87.80 & 85.70 & 62.84 & 67.24 & 41.90 & 45.51 & 65.46 & 67.96 \\
\midrule
CLIP & L/14 & 68.20 & 56.00 & 84.20 & 83.50 & 65.70 & 63.16 & 36.98 & 35.06 & 63.77 & 59.43 \\
ERCLIP M & L/14 & 78.30 & 81.30 & 89.00 & 88.00 & 71.62 & 75.28 & 47.59 & 51.12 & 71.63 & 73.92 \\
\rowcolor{gray!12}ERCLIP M+P & L/14 & 76.40 & 79.90 & 89.50 & 89.30 & 72.14 & 74.02 & 48.46 & 51.10 & 71.62 & 73.58 \\
\rowcolor{gray!12}ERCLIP RT+P & L/14 & 75.40 & 80.00 & 90.40 & 89.70 & 71.82 & 73.54 & 47.91 & 51.17 & 71.38 & 73.60 \\
\bottomrule
\end{tabular}
}%
\caption{Long-caption retrieval (Recall@1, \%). CLIP is the pretrained model without fine-tuning. ERCLIP rows are trained for 1 epoch on 20k pairs: M is mean pooling, M+P adds $\mathcal{L}_{\mathrm{pres}}$, and RT+P is residual Transformer fusion with $\mathcal{L}_{\mathrm{pres}}$ (shaded). Grey rows are published results of methods trained on larger long-caption corpora and are given for reference only. ``--'': not available.}
\label{tab:long_text_retrieval}\vspace{-2em}
\end{table}

\subsection{Long-Caption Retrieval}
Table~\ref{tab:long_text_retrieval} shows that one epoch on 20k pairs raises CLIP's four-benchmark average by 6.5 (B/32), 7.6 (B/16), and 11.2 points (L/14), mostly in text-to-image retrieval. Preservation does not trade retrieval away: M+P gains 1.0 point over M on B/32, M+P and RT+P gain 0.4 and 0.3 points on B/16, and on L/14 all three are within 0.3 points. The residual Transformer does not surpass mean pooling, as expected from a gate that departs from the mean only when the data support it. Methods trained on large corpora (grey rows) remain clearly stronger: SmartCLIP leads the best ERCLIP variant by 13.1 (B/16) and 10.2 (L/14) average points, and LongCLIP by 4.7 and 3.4. They differ in data (1M+ vs.\ 20k captions), compute, and test-time context (248+ vs.\ 77 tokens), and none has been retrained on our 20k pairs, so these rows are a reference rather than a same-budget baseline. One epoch on 20k pairs closes 62\% (B/16) and 77\% (L/14) of the CLIP-to-LongCLIP gap. Much of the alignment is thus reachable with little data, though this does not show greater data efficiency.

\begin{table}[t]
\centering
\setlength{\tabcolsep}{3pt}
\resizebox{.8\linewidth}{!}{%
\begin{tabular}{llc rr rr rrr}
\toprule
\textbf{ViT} & \textbf{Model} & $\mathcal{L}_{\mathrm{pres}}$ & \multicolumn{2}{c}{\textbf{Long avg.}} & \multicolumn{2}{c}{\textbf{COCO}} & \multicolumn{3}{c}{\textbf{Zero-shot cls.}} \\
\cmidrule(lr){4-5}\cmidrule(lr){6-7}\cmidrule(lr){8-10}
 & & & I2T & T2I & I2T & T2I & C10 & C100 & IN \\
\midrule
\multirow{4}{*}{B/32} & CLIP (no fine-tuning) & -- & 59.33 & 51.93 & 50.00 & 30.37 & 89.76 & 64.31 & 63.34 \\
 & ERCLIP-v1 LSTM & $\times$ & 45.99 & 44.23 & 45.94 & 34.32 & 86.42 & 56.60 & 49.48 \\
 & ERCLIP-v1 Transformer & $\times$ & 60.33 & 62.32 & 53.34 & 39.31 & 88.13 & 64.89 & 58.04 \\
\cmidrule(lr){2-10}
 & ERCLIP M & $\times$ & 60.57 & 61.63 & 54.40 & 39.16 & 87.75 & 62.45 & 57.51 \\
\rowcolor{gray!12}\cellcolor{white} & ERCLIP M+P & $\checkmark$ & 61.38 & 62.88 & 56.40 & 39.74 & 87.51 & 63.42 & 59.64 \\
\rowcolor{gray!12}\cellcolor{white} & ERCLIP RT+P & $\checkmark$ & 61.22 & 62.31 & 55.46 & 39.68 & 86.49 & 63.66 & 59.43 \\
\midrule
\multirow{4}{*}{B/16} & CLIP (no fine-tuning) & -- & 62.42 & 55.93 & 51.72 & 32.66 & 90.81 & 67.22 & 68.35 \\
 & ERCLIP-v1 LSTM & $\times$ & 51.27 & 51.17 & 50.68 & 38.38 & 85.51 & 59.87 & 54.09 \\
 & ERCLIP-v1 Transformer & $\times$ & 65.94 & 67.68 & 59.98 & 44.46 & 89.22 & 67.66 & 62.57 \\
\cmidrule(lr){2-10}
 & ERCLIP M & $\times$ & 65.17 & 67.61 & 59.00 & 44.28 & 91.61 & 67.92 & 62.30 \\
\rowcolor{gray!12}\cellcolor{white} & ERCLIP M+P & $\checkmark$ & 65.23 & 68.38 & 61.36 & 44.51 & 91.83 & 68.73 & 64.09 \\
\rowcolor{gray!12}\cellcolor{white} & ERCLIP RT+P & $\checkmark$ & 65.46 & 67.96 & 61.34 & 44.36 & 89.34 & 68.02 & 64.15 \\
\midrule
\multirow{4}{*}{L/14} & CLIP (no fine-tuning) & -- & 63.77 & 59.43 & 56.10 & 35.33 & 95.70 & 76.44 & 75.53 \\
 & ERCLIP-v1 LSTM & $\times$ & 57.34 & 59.34 & 55.40 & 44.53 & 93.60 & 74.34 & 63.13 \\
 & ERCLIP-v1 Transformer & $\times$ & 70.94 & 73.91 & 65.62 & 50.35 & 94.86 & 77.02 & 72.30 \\
\cmidrule(lr){2-10}
 & ERCLIP M & $\times$ & 71.63 & 73.92 & 65.24 & 50.55 & 94.64 & 76.94 & 71.76 \\
\rowcolor{gray!12}\cellcolor{white} & ERCLIP M+P & $\checkmark$ & 71.62 & 73.58 & 66.58 & 49.95 & 95.37 & 78.29 & 73.36 \\
\rowcolor{gray!12}\cellcolor{white} & ERCLIP RT+P & $\checkmark$ & 71.38 & 73.60 & 66.36 & 49.96 & 95.92 & 78.83 & 73.26 \\
\bottomrule
\end{tabular}
}%
\caption{Long-caption retrieval versus the tasks CLIP already solves. \emph{Long avg.}\ is Recall@1 averaged over the four long-caption benchmarks, COCO is 5k-test Recall@1, and classification is zero-shot accuracy on CIFAR-10/100 (C10/C100) and ImageNet (IN). The lower block of each backbone is the matched 1-epoch comparison.}
\label{tab:basic_tasks}\vspace{-2em}
\end{table}

\subsection{Effect on Basic Tasks}
\label{sec:basic}
Table~\ref{tab:basic_tasks} places long-caption retrieval next to the tasks CLIP already solves. ERCLIP-v1 shows the zero-shot degradation known for classification fine-tuning \cite{wortsman2022robust} under a retrieval objective. Mean and Transformer fusion cost 3.2--6.1 ImageNet points on every backbone while COCO retrieval improves, as expected since the short captions come from COCO. The LSTM loses 12.4--14.3 ImageNet and 2.1--7.7 CIFAR-100 points and has the worst long-caption retrieval. This ordering follows how far each fusion output may move from the CLIP space: mean pooling and the scorer combine CLIP sentence embeddings, while the LSTM emits a new vector. We do not measure drift directly, so this is a pattern consistent across backbones, not a measured relation.
\begin{figure}[!ht]
    \centering
    \begin{minipage}[b]{0.5\linewidth}
        \centering
\resizebox{.8\linewidth}{!}{%
\begin{tabular}{ll rr rr rr}
\toprule
 & & \multicolumn{2}{c}{\textbf{Long avg.}} & \multicolumn{2}{c}{\textbf{COCO}} & \multicolumn{2}{c}{\textbf{Zero-shot}} \\
\cmidrule(lr){3-4}\cmidrule(lr){5-6}\cmidrule(lr){7-8}
 & & I2T & T2I & I2T & T2I & C10 & IN \\
\midrule
\multirow{5}{*}{$\lambda_p$} & 0.0 & 60.36 & 61.44 & 53.96 & 39.07 & 88.72 & 58.16 \\
 & 0.25 & 61.45 & 62.32 & 54.78 & 39.42 & 88.71 & 59.29 \\
 & 0.5 & 61.75 & 62.71 & 55.74 & 39.61 & 88.87 & 59.67 \\
 & 1.0 & 61.60 & 62.49 & 56.20 & 39.48 & 88.52 & 60.06 \\
 & 2.0 & 61.56 & 62.33 & 56.10 & 39.20 & 88.65 & 60.76 \\
\midrule
\multirow{5}{*}{$N$} & 1k & 60.08 & 52.50 & 51.22 & 30.45 & 89.92 & 63.48 \\
 & 5k & 59.81 & 58.85 & 53.00 & 36.45 & 90.19 & 61.08 \\
 & 10k & 62.37 & 61.52 & 55.22 & 38.14 & 89.00 & 60.91 \\
 & 20k & 61.60 & 62.49 & 56.20 & 39.48 & 88.52 & 60.06 \\
 & 50k (all) & 61.43 & 63.12 & 57.42 & 41.19 & 88.10 & 59.72 \\
\bottomrule
\end{tabular}
}\captionof{table}{Ablations on ViT-B/32 (mean pooling, 1 epoch): preservation weight $\lambda_p$ with $N{=}20$k pairs, and training-set size $N$ with $\lambda_p{=}1$. These runs fully decay the learning rate within the epoch and are therefore not directly comparable to Table~\ref{tab:basic_tasks}. Long avg.\ is Recall@1 averaged over the four long-caption benchmarks; IN is ImageNet zero-shot accuracy.}
\label{tab:ablation}
    \end{minipage}%
    \hfill
    \begin{minipage}[b]{0.4\linewidth}
        \centering
        \includegraphics[width=\linewidth]{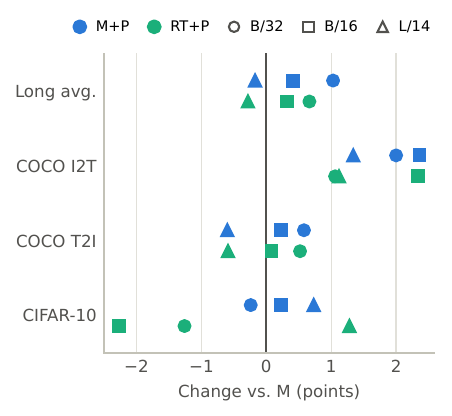}
        \caption{Change of M+P and RT+P relative to M (epoch 1). Colour: fusion; marker: backbone.}
        \label{fig:preservation}
    \end{minipage}%
    \vspace{-1em}
\end{figure}

\begin{figure}[t]
\centering
\includegraphics[width=.8\linewidth]{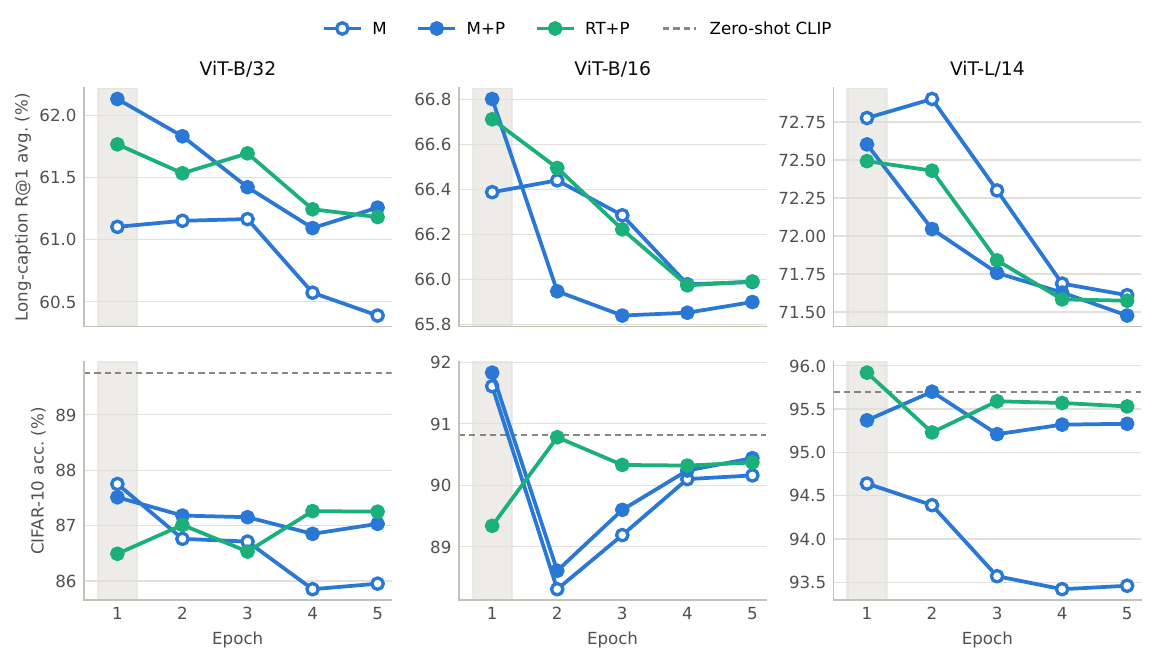}
\caption{Long-caption R@1 (four-benchmark average, top) and zero-shot CIFAR-10 accuracy (bottom) over five epochs. Each panel uses its own scale. The shaded epoch is the budget reported in all tables, and the dashed line is the pretrained model. Colour denotes the fusion type; hollow markers are runs without preservation.}
\label{fig:epochs}\vspace{-1em}
\end{figure}

Fig.~\ref{fig:preservation} isolates the matched rows of Table~\ref{tab:basic_tasks}. Preservation acts most clearly on short-caption alignment: COCO I2T rises over M in every run, by 1.3--2.4 points with M+P and 1.1--2.3 with RT+P, while long-caption R@1 stays within $-0.3$ to $+1.0$ points of M and COCO T2I within 0.6. Anchoring thus restores the short-caption interface without competing with the long-caption loss. The fusion module matters mainly for classification. With mean pooling, CIFAR-10 stays within $-0.2$ to $+0.7$ points of M. RT+P is 1.3--2.3 points below M on B/32 and B/16 at epoch 1 but reaches M+P by epoch 2 (Fig.~\ref{fig:epochs}), suggesting that the new scorer briefly perturbs the shared encoder before its gate settles; on L/14 it is 1.3 points above M. Under a one-epoch budget M+P is therefore the safest variant. On ImageNet, the hardest zero-shot test, one epoch without preservation costs 3.8 (L/14) to 6.1 (B/16) points, and $\mathcal{L}_{\mathrm{pres}}$ recovers 1.5--2.1 of them on every backbone and fusion type (30--42\%); CIFAR-100 improves over M in every run. Preservation reduces the erosion but does not remove it, and it is not an advantage over full-scale training: LongCLIP reports zero-shot accuracy on par with or above ours while being stronger on long captions \cite{zhang2024longclip}. Its role is to make small-data adaptation safe.

\subsection{Training Length}
\label{sec:length}
Longer training does not help (Fig.~\ref{fig:epochs}). From epoch 1 to 5, long-caption R@1 falls by 0.4--1.2 points for every variant while COCO R@1 rises by 0.5--1.2 points on the B backbones and is flat on L/14: the model drifts toward the COCO caption style. Without preservation CIFAR-10 also falls, by 1.2--1.8 points; with it the B/32 decline shrinks to 0.5. Preservation limits this drift but not the loss of long-caption alignment, so a fixed early budget is needed. We have not tested whether the decline persists with a learned temperature, larger batches, or a schedule matched to a longer budget, so one epoch is an empirical choice for our configuration.

\subsection{Ablations}
\noindent\textbf{Components.} In the matched block of Table~\ref{tab:basic_tasks}, CLIP$\to$M adds sentence-level training with $\mathcal{L}_{\mathrm{long}}$, M$\to$M+P adds only $\mathcal{L}_{\mathrm{pres}}$, and M+P$\to$RT+P changes only the fusion. Not included are RT without $\mathcal{L}_{\mathrm{pres}}$ and an $\mathcal{L}_{\mathrm{short}}$-only run, which would separate the long-caption loss from fine-tuning on COCO captions; the v1 rows are not matched and show only the direction of the fusion effect.

\noindent\textbf{Preservation weight and data size.} Table~\ref{tab:ablation} uses a one-epoch schedule that fully decays the learning rate, so its rows compare only with each other. Raising $\lambda_p$ from 0 to 2 raises ImageNet monotonically (58.2 to 60.8), while long-caption and COCO R@1 rise by up to 1.3 points and plateau for $\lambda_p\geq0.5$ and CIFAR-10 stays within 0.4 points. We use $\lambda_p{=}1$, untuned on the test sets. Long-caption R@1 saturates early (56.3 at 1k, 61.9 at 10k, within 0.4 to 50k) while COCO R@1 rises (40.8 to 49.3) and ImageNet falls (63.5 to 59.7): beyond about 10k pairs extra data mainly shifts the model toward COCO captions. We use $N{=}20$k.

\section{Discussion and Conclusion}
With 20k pairs and one epoch, ERCLIP gains 6.5--11.2 points of long-caption R@1 without changing CLIP's inference, and preservation recovers 30--42\% of the ImageNet loss. The fusion choice, not only the objective, sets the cost of adaptation.

\noindent\textbf{When to use it.} With a million long captions and several GPUs, LongCLIP or SmartCLIP is the better choice: they read the full caption and retrieve better. ERCLIP fits when only thousands of long captions and one GPU are available, e.g., in a specialised domain, and the result must remain a standard CLIP checkpoint that keeps short-caption retrieval and zero-shot classification usable. Its inference cost equals CLIP's, a small advantage over LongCLIP, whose extra cost is confined to a 248-token text pass; we did not benchmark it.

\noindent\textbf{Limitations.} Test captions are truncated to 77 tokens and sentence-wise inference is not evaluated; no baseline is retrained on our 20k pairs; results are single-seed; the one-epoch budget holds for one optimisation setting; the link between drift and zero-shot loss is inferred rather than measured; and all data are general-domain.

\section*{Acknowledgement}

This publication has emanated from research supported in part by research grants from Science Foundation Ireland under grant numbers SFI/13/RC/2106\_P2 and 18/CRT/6223, and co-funded by the European Regional Development Fund.

\bibliographystyle{splncs04}
\bibliography{myref}

@inproceedings{li2023blip2,
  title={Blip-2: Bootstrapping language-image pre-training with frozen image encoders and large language models},
  author={Li, Junnan and Li, Dongxu and Savarese, Silvio and Hoi, Steven},
  booktitle={International conference on machine learning},
  pages={19730--19742},
  year={2023},
  organization={PMLR}
}

@misc{lin2015microsoftcococommonobjects,
      title={Microsoft COCO: Common Objects in Context}, 
      author={Tsung-Yi Lin and Michael Maire and Serge Belongie and Lubomir Bourdev and Ross Girshick and James Hays and Pietro Perona and Deva Ramanan and C. Lawrence Zitnick and Piotr Dollár},
      year={2015},
      eprint={1405.0312},
      archivePrefix={arXiv},
      primaryClass={cs.CV},
      url={https://arxiv.org/abs/1405.0312}, 
}

@inproceedings{li2022blip,
      title={BLIP: Bootstrapping Language-Image Pre-training for Unified Vision-Language Understanding and Generation}, 
      author={Junnan Li and Dongxu Li and Caiming Xiong and Steven Hoi},
      year={2022},
      booktitle={ICML},
}

@inproceedings{radford2021learning,
  title={Learning transferable visual models from natural language supervision},
  author={Radford, Alec and Kim, Jong Wook and Hallacy, Chris and Ramesh, Aditya and Goh, Gabriel and Agarwal, Sandhini and Sastry, Girish and Askell, Amanda and Mishkin, Pamela and Clark, Jack and others},
  booktitle={International conference on machine learning},
  pages={8748--8763},
  year={2021},
  organization={PMLR}
}

@article{loshchilov2017adamw,
  title={Decoupled weight decay regularization},
  author={Loshchilov, Ilya and Hutter, Frank},
  journal={arXiv preprint arXiv:1711.05101},
  year={2017}
}

@article{raffel2020t5,
  title={Exploring the limits of transfer learning with a unified text-to-text transformer},
  author={Raffel, Colin and Shazeer, Noam and Roberts, Adam and Lee, Katherine and Narang, Sharan and Matena, Michael and Zhou, Yanqi and Li, Wei and Liu, Peter J},
  journal={Journal of machine learning research},
  volume={21},
  number={140},
  pages={1--67},
  year={2020}
}

@inproceedings{zhang2024longclip,
  title={Long-clip: Unlocking the long-text capability of clip},
  author={Zhang, Beichen and Zhang, Pan and Dong, Xiaoyi and Zang, Yuhang and Wang, Jiaqi},
  booktitle={European conference on computer vision},
  pages={310--325},
  year={2024},
  organization={Springer}
}

@inproceedings{jia2021align,
  title={Scaling up visual and vision-language representation learning with noisy text supervision},
  author={Jia, Chao and Yang, Yinfei and Xia, Ye and Chen, Yi-Ting and Parekh, Zarana and Pham, Hieu and Le, Quoc and Sung, Yun-Hsuan and Li, Zhen and Duerig, Tom},
  booktitle={International conference on machine learning},
  pages={4904--4916},
  year={2021},
  organization={PMLR}
}

@inproceedings{zhai2023sigmoid,
  title={Sigmoid loss for language image pre-training},
  author={Zhai, Xiaohua and Mustafa, Basil and Kolesnikov, Alexander and Beyer, Lucas},
  booktitle={Proceedings of the IEEE/CVF international conference on computer vision},
  pages={11975--11986},
  year={2023}
}

@misc{krizhevsky2009cifar,
  title={Learning multiple layers of features from tiny images.(2009)},
  author={Krizhevsky, Alex and Hinton, Geoffrey and others},
  year={2009}
}

@article{russakovsky2015imagenet,
  title={Imagenet large scale visual recognition challenge},
  author={Russakovsky, Olga and Deng, Jia and Su, Hao and Krause, Jonathan and Satheesh, Sanjeev and Ma, Sean and Huang, Zhiheng and Karpathy, Andrej and Khosla, Aditya and Bernstein, Michael and others},
  journal={International journal of computer vision},
  volume={115},
  number={3},
  pages={211--252},
  year={2015},
  publisher={Springer}
}

@inproceedings{xie2025smartclip,
  title={SmartCLIP: Modular Vision-language Alignment with Identification Guarantees},
  author={Xie, Shaoan and Lingjing, Lingjing and Zheng, Yujia and Yao, Yu and Tang, Zeyu and Xing, Eric P and Chen, Guangyi and Zhang, Kun},
  booktitle={Proceedings of the Computer Vision and Pattern Recognition Conference},
  pages={29780--29790},
  year={2025}
}

@inproceedings{chen2024sharegpt4v,
  title={Sharegpt4v: Improving large multi-modal models with better captions},
  author={Chen, Lin and Li, Jinsong and Dong, Xiaoyi and Zhang, Pan and He, Conghui and Wang, Jiaqi and Zhao, Feng and Lin, Dahua},
  booktitle={European Conference on Computer Vision},
  pages={370--387},
  year={2024},
  organization={Springer}
}

@inproceedings{wu2026himoclip,
  title={HiMo-CLIP: Modeling Semantic Hierarchy and Monotonicity in Vision-Language Alignment},
  author={Wu, Ruijia and Chen, Ping and Shen, Fei and Zhao, Shaoan and Hui, Qiang and Gao, Huanlin and Lu, Ting and Liu, Zhaoxiang and Zhao, Fang and Wang, Kai and others},
  booktitle={Proceedings of the AAAI Conference on Artificial Intelligence},
  volume={40},
  number={32},
  pages={26974--26982},
  year={2026}
}

@article{hochreiter1997lstm,
  title={Long short-term memory},
  author={Hochreiter, Sepp and Schmidhuber, J{\"u}rgen},
  journal={Neural computation},
  volume={9},
  number={8},
  pages={1735--1780},
  year={1997},
  publisher={MIT press}
}

@inproceedings{onoe2024docci,
  author        = {Yasumasa Onoe and Sunayana Rane and Zachary Berger and Yonatan Bitton and Jaemin Cho and Roopal Garg and
    Alexander Ku and Zarana Parekh and Jordi Pont-Tuset and Garrett Tanzer and Su Wang and Jason Baldridge},
  title         = {{DOCCI: Descriptions of Connected and Contrasting Images}},
  booktitle     = {ECCV},
  year          = {2024}
}

@inproceedings{najdenkoska2025tulip,
title={{TULIP}: Token-length Upgraded {CLIP}},
author={Ivona Najdenkoska and Mohammad Mahdi Derakhshani and Yuki M Asano and Nanne Van Noord and Marcel Worring and Cees G. M. Snoek},
booktitle={The Thirteenth International Conference on Learning Representations},
year={2025},
url={https://openreview.net/forum?id=r9oqHOdoHf}
}

@InProceedings{Urbanek_2024_CVPR,
    author    = {Urbanek, Jack and Bordes, Florian and Astolfi, Pietro and Williamson, Mary and Sharma, Vasu and Romero-Soriano, Adriana},
    title     = {A Picture is Worth More Than 77 Text Tokens: Evaluating CLIP-Style Models on Dense Captions},
    booktitle = {Proceedings of the IEEE/CVF Conference on Computer Vision and Pattern Recognition (CVPR)},
    month     = {June},
    year      = {2024},
    pages     = {26700-26709}
}

@inproceedings{li2023declip,
  title={Supervision Exists Everywhere: A Data Efficient Contrastive Language-Image Pre-training Paradigm},
  author={Li, Yangguang and Liang, Feng and Zhao, Lichen and Cui, Yufeng and Ouyang, Wanli and Shao, Jing and Yu, Fengwei and Yan, Junjie},
  booktitle={International Conference on Learning Representations}
}

@inproceedings{yang2023alip,
  title     = {{ALIP: Adaptive Language-Image Pre-Training with Synthetic Caption}},
  author    = {Yang, Kaicheng and Deng, Jiankang and An, Xiang and Li, Jiawei and Feng, Ziyong and Guo, Jia and Yang, Jing and Liu, Tongliang},
  booktitle = {International Conference on Computer Vision},
  year      = {2023},
  pages     = {2922-2931},
  doi       = {10.1109/ICCV51070.2023.00273},
  url       = {https://mlanthology.org/iccv/2023/yang2023iccv-alip/}
}

@inproceedings{wortsman2022robust,
  title={Robust fine-tuning of zero-shot models},
  author={Wortsman, Mitchell and Ilharco, Gabriel and Kim, Jong Wook and Li, Mike and Kornblith, Simon and Roelofs, Rebecca and Lopes, Raphael Gontijo and Hajishirzi, Hannaneh and Farhadi, Ali and Namkoong, Hongseok and Schmidt, Ludwig},
  booktitle={Proceedings of the IEEE/CVF Conference on Computer Vision and Pattern Recognition (CVPR)},
  pages={7959--7971},
  year={2022}
}

@inproceedings{kumar2022finetuning,
  title={Fine-Tuning can Distort Pretrained Features and Underperform Out-of-Distribution},
  author={Kumar, Ananya and Raghunathan, Aditi and Jones, Robbie and Ma, Tengyu and Liang, Percy},
  booktitle={International Conference on Learning Representations},
  year={2022}
}

@inproceedings{goyal2023finetune,
  title={Finetune like you pretrain: Improved finetuning of zero-shot vision models},
  author={Goyal, Sachin and Kumar, Ananya and Garg, Sankalp and Kolter, Zico and Raghunathan, Aditi},
  booktitle={Proceedings of the IEEE/CVF Conference on Computer Vision and Pattern Recognition (CVPR)},
  pages={19338--19347},
  year={2023}
}

@article{li2018learning,
  title={Learning without forgetting},
  author={Li, Zhizhong and Hoiem, Derek},
  journal={IEEE Transactions on Pattern Analysis and Machine Intelligence},
  volume={40},
  number={12},
  pages={2935--2947},
  year={2018},
  publisher={IEEE}
}

@article{kirkpatrick2017overcoming,
  title={Overcoming catastrophic forgetting in neural networks},
  author={Kirkpatrick, James and Pascanu, Razvan and Rabinowitz, Neil and Veness, Joel and Desjardins, Guillaume and Rusu, Andrei A and Milan, Kieran and Quan, John and Ramalho, Tiago and Grabska-Barwinska, Agnieszka and others},
  journal={Proceedings of the National Academy of Sciences},
  volume={114},
  number={13},
  pages={3521--3526},
  year={2017},
  publisher={National Academy of Sciences}
}
\end{document}